\documentstyle[11pt,epsf]{article}

\newskip\humongous \humongous=0pt plus 1000pt minus 100pt
\def\caja{\mathsurround=0pt}
\def\eqalign#1{\,\vcenter{\openup1\jot \caja
       \ialign{\strut \hfil$\displaystyle{##}$&$
        \displaystyle{{}##}$\hfil\crcr#1\crcr}}\,}
\newif\ifdtup

\newcounter{eqnumber}
\renewcommand{\theeqnumber}{\arabic{eqnumber}}
\def\equn{
\refstepcounter{eqnumber}
\eqno({\rm \theeqnumber})
}

\def\eqn#1{eq.~(\ref{#1})}

\def\fig#1{fig.~{\ref{#1}}}

\newbox\charbox
\newbox\slabox
\def\s#1{{      
        \setbox\charbox=\hbox{$#1$}
        \setbox\slabox=\hbox{$/$}
        \dimen\charbox=\ht\slabox
        \advance\dimen\charbox by -\dp\slabox
        \advance\dimen\charbox by -\ht\charbox
        \advance\dimen\charbox by \dp\charbox
        \divide\dimen\charbox by 2
        \raise-\dimen\charbox\hbox to \wd\charbox{\hss/\hss}
        \llap{$#1$}
}}

\def\spa#1.#2{\left\langle#1\,#2\right\rangle}
\def\spb#1.#2{\left[#1\,#2\right]}
\def\spab#1.#2.#3{\langle\mskip-1mu{#1} 
                  | #2 | {#3}\mskip-1mu\rangle}
\def\spba#1.#2.#3{\langle\mskip-1mu{#1}^+ 
                  | #2 | {#3}^+\mskip-1mu\rangle}
\def\lor#1.#2{\left(#1\,#2\right)}

\catcode`@=11  

\def\Tr{\, {\rm Tr}}

\def\eps{\epsilon}
\def\Ord{{\cal O}}

\def\pol{\varepsilon}

\def\la{\langle}
\def\ra{\rangle}
\def\tree{{\rm tree}}

\def\lsl{\not{\hbox{\kern-2.3pt $\ell$}}}
\def\ksl{\not{\hbox{\kern-2.3pt $k$}}}
\def\Ksl{\not{\hbox{\kern-2.6pt $K$}}}

\def\Soft{\mathop {\cal S} \nolimits}

\def\Psl{\not{\hbox{\kern-2.6pt $P$}}}

\begin{document}

\begin{titlepage}

\begin{flushright}
hep-th/9904026 \hfill UCLA/99/TEP/10\\
HUTP-99/A017\\
April, 1999\\
\end{flushright}

\vskip 2.cm

\begin{center}
\begin{Large}
{\bf Perturbative Gravity from QCD Amplitudes}
\end{Large}

\vskip 2.cm

{\large Z. Bern$^\star$}

\vskip 0.2cm

{\it Department of Physics,\\
University of California at Los Angeles,\\
Los Angeles,  CA 90095-1547}

\vskip .4cm
and 
\vskip .3 cm 

{\large A. K. Grant$^\dagger$}

\vskip 0.2cm

{\it Department of Physics,\\
Harvard University,\\
Cambridge, MA 02138}

\vskip 1cm
\end{center}

\begin{abstract}
We demonstrate that QCD gluon amplitudes can be used to construct a
Lagrangian for gravity. This procedure makes use of perturbative
`squaring' relations between gravity and gauge theory that follow from
string theory.  We explicitly carry out the construction for up to
five-point interactions and discuss a set of field variables in the
Einstein-Hilbert Lagrangian for interpreting the Lagrangian obtained
from QCD.  A spin-off from our analysis is that it can be used to
provide simpler tree-level gravity Feynman rules than for
conventional gauges.
\end{abstract}


\vfill
\noindent\hrule width 3.6in\hfil\break
${}^{\star}$ Research supported by the US Department of Energy
under grant DE-FG03-91ER40662.\\
$^\dagger$ Research supported by the National Science Foundation
grant  PHY-9802709.
\hfil\break

\end{titlepage}

\baselineskip 16pt


\section{Introduction}
\label{IntroSection}

Although both gravity and gauge theories contain a local symmetry,
they have rather disparate physical properties.  QCD, for example, is
a confining theory while gravity is not. Similarly, in four dimensions
QCD is renormalizable while field theories of gravity are
non-renormalizable.  Nonetheless, within the context of the
perturbative expansions some remarkable `squaring' relations exist
between the tree-level $S$-matrix elements of gravity and gauge
theories in both string and field
theories~\cite{KLT,BGK,BDDPR,AllPlusGravity,MHVGravity}.  These
squaring relations imply that gauge symmetry is more closely related
to the diffeomorphism invariance of gravity than one might suspect
based on the respective Lagrangians.  Recently, there has also been a
resurgence of interest in perturbative gravity both in anti-de Sitter
spaces~\cite{AdS} and in phenomenological applications of large
compact dimensions~\cite{Dimopoulos}, further motivating 
an investigation of these relations.

In this letter we will show that one can construct a low energy
Lagrangian for gravity directly from QCD $S$-matrix elements and then
discuss how this Lagrangian relates to the usual Einstein-Hilbert
Lagrangian.  The starting point of our investigation is the Kawai,
Lewellen and Tye (KLT) relations expressing closed string tree
amplitudes in terms of open string amplitudes~\cite{KLT}.  These
relations arise from the property that the integrand of a closed
string amplitude is composed of left- and right-mover open string
components (see e.g.~ref.~\cite{GSW}).  In the low energy limit where
string theory reduces to field theory, the KLT relations express
gravity tree amplitudes in terms of a sum of products of `left' gauge
theory amplitudes and `right' gauge theory amplitudes. 

The KLT relations were first exploited by Berends, Giele and Kuijf to
obtain an infinite sequence of maximally helicity violating pure
gravity tree amplitudes~\cite{BGK} using known gauge theory
results. By unitarity, tree-level relations necessarily imply that
loop amplitudes in the two theories are also related.  Indeed, the
method of constructing $S$-matrix elements via their analytic
properties (see refs.~\cite{Review}) provides a means for
exploiting this to obtain loop-level (super) gravity amplitudes.
Using these ideas, the divergence structure of $N=8$ supergravity has
been investigated with the result that it appears to be less divergent
than previously thought~\cite{BDDPR}.  Two infinite sequences of
maximally helicity violating one-loop amplitudes in gravity theories
have also been constructed~\cite{MHVGravity}.  These
may be viewed as results in an effective field theory of gravity
which necessarily must match the low energy limit of a more
fundamental theory, such as string or $M$ theory.  The ability to use
the squaring relation between gravity and gauge theory to perform
non-trivial gravity computations suggests that one can develop a
deeper understanding of perturbative gravity by exploring this
relationship.

{}From the point of view of the field theory Lagrangians, the KLT
relations are rather mysterious: the Einstein-Hilbert Lagrangian does
not factorize in any obvious way in terms of the Yang-Mills
Lagrangian.  It is not even completely clear what the notion of `left' and
`right' parts of the theory mean given that the graviton is a
symmetric tensor. Another obvious difficulty is that the Yang-Mills
Lagrangian has only three- and four-point interactions while the
Einstein-Hilbert Lagrangian has an infinite set of interactions.

Nevertheless, the gravity $S$-matrix does have the property that it is
composed of products of two gauge theory amplitudes and that a
graviton, very roughly speaking, is composed of `left' and `right'
gauge fields, i.e., $h_{\mu\nu} \sim {\bar A}_\mu A_\nu$.  When one
has a property of the $S$-matrix, a natural idea is to organize the
underlying formalism so that it reflects that property.  As a well
known example, the general superiority of Feynman diagrams 
as compared to time ordered perturbation theory follows from their
preservation of manifest Lorentz symmetry.  Here we wish to rearrange
the gravity Lagrangian so it captures properties associated with the
factorization of the $S$-matrix into sums of products of gauge theory
amplitudes.

The analysis of this letter will consist of two parts.  In the first
part we will systematically construct a gravity Lagrangian from QCD
gluon amplitudes using the KLT relations.  We explicitly carry out the
procedure through five graviton interactions; in principle, the
procedure can be carried out to arbitrarily high orders, although as a
practical matter the complexity of the computations increases rapidly
with the number of gravitons.  Nevertheless, the procedure shows that
a Lagrangian for gravity can be obtained using QCD amplitudes. The
gravity Lagrangian obtained in this way has the properties that (a) by
construction it produces the correct tree-level amplitudes and (b)
graviton Lorentz indices associated with the `left' gauge theory do
not contract with ones associated with the `right' gauge theory. A
Lagrangian with the property that the associated Feynman diagrams
factorize into left and right parts has been presented previously by
Siegel~\cite{Siegel}. A spin-off from our analysis is that it can be
used to provide a simpler set of gravity Feynman rules than the
conventional~\cite{DeWitt} de Donder gauge rules.  Moreover, in our
construction, the gravity three vertex is given directly as sums of
squares of gauge theory vertices.

In the second part, we will find a set of field variables which allows
us to interpret the gravity Lagrangian obtained from QCD in terms of
the conventional Einstein-Hilbert Lagrangian.  Non-linear field
redefinitions and gauge fixings have been used previously by van de
Ven~\cite{vandeVen} to simplify the computation of two-loop
divergences in gravity.  Here we wish to consider such field redefinitions
and gauge fixings in order to find a particular set of
field variables which helps clarify the connection to QCD.  This
construction indicates that there is a closer correspondence between
ordinary gauge invariance and the diffeomorphism invariance of
Einstein gravity than one might have suspected.

\section{Review of the KLT relations and applications}

A basic property of integrands for closed string amplitudes is that
they factorize into open string integrands (except
for the momentum zero mode). At tree-level, KLT~\cite{KLT} used this
property to find simple relationships expressing closed string
amplitudes in terms of products of open string amplitudes.  In the
infinite string tension limit where string theory reduces to an
effective field theory, for the four- and five-point amplitudes these
relations are,
$$
\eqalign{
{\cal M}_4^{\rm tree} (1,2,3,4) & =
     - i \Bigl({\kappa \over 2} \Bigr)^2
    s_{12} A_4^{\rm tree} (1,2,3,4) \, A_4^{\rm tree}(1,2,4,3)\,, \cr
{\cal M}_5^{\rm tree}
(1,2,3,4,5) & = i \Bigl({\kappa\over 2} \Bigr)^3 \Bigl[ 
     s_{12} s_{34}  A_5^{\rm tree}(1,2,3,4,5)
                                     A_5^{\rm tree}(2,1,4,3,5)  \cr
& \hskip 1.5 cm 
      + s_{13}s_{24} A_5^{\rm tree}(1,3,2,4,5) \, 
                           A_5^{\rm tree}(3,1,4,2,5) \Bigr]\,, \cr}
\equn\label{KLTExamples}
$$
where, $\kappa^2 = 32 \pi G_N$, $s_{ij} = (k_i + k_j)^2$, the $A_n$
are color-ordered gauge theory partial amplitudes~\cite{ManganoReview}
and the ${\cal M}_n$ are gravity amplitudes. Color does not appear in
these relations because the gauge theory partial amplitudes are
independent of color.  The arguments of the amplitudes label the
external legs. These relations hold for any particle states appearing
in any closed string theory since they follow from the basic
factorization of the string integrands into `left' and `right'
sectors.  In particular, they hold in pure Einstein gravity, gravity
coupled to a dilaton, anti-symmetric tensor or gauge field, and in
supergravity theories.  Moreover, these KLT relations generalize to an
arbitrary number of external legs. (Explicit expressions may be found
in Appendix~A of ref.~\cite{MHVGravity}.) A basic property of the KLT
formulas is that any Lorentz indices associated with the `left' gauge
theory amplitude do not contract with the Lorentz indices of the
`right' gauge theory amplitude.

The KLT relations give tree-level gravity amplitudes directly from
known gauge theory amplitudes, which are generally far easier to
compute.  This was first applied by Berends, Giele and
Kuijf~\cite{BGK} to obtain an infinite sequence of maximally helicity
violating $n$-point Einstein gravity tree amplitudes using known
QCD gluon amplitudes.  Beyond tree-level, one may exploit the KLT
relations using the cutting method developed for performing QCD amplitude
calculations of phenomenological interest~\cite{Review}.  With this
method one can reconstruct complete loop amplitudes (in a given
dimensional regularization scheme) from their $D$-dimensional
unitarity cuts.  The one-loop amplitudes are specified in terms of
tree amplitudes which satisfy the KLT relations; one can then obtain
higher loop amplitudes by iterating the procedure.  In this way, one
can obtain loop-level (super) gravity $S$-matrix elements without
reference to a Lagrangian or Feynman
rules~\cite{BDDPR,AllPlusGravity,MHVGravity}.  The efficiency of the
computational method follows from the fact that one is {\it recycling}
previously performed gauge theory calculations to obtain new
results in gravity theories.

%
\begin{figure}[ht]
\centerline{\epsfxsize 1.9 truein \epsfbox{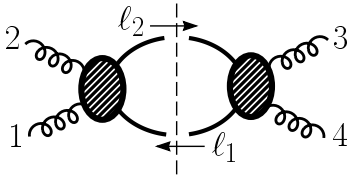}}
\vskip -.3 cm
\caption[]{
\label{TwoParticleFigure}
\small One can recycle tree amplitudes into loop amplitudes via
$D$-dimensional unitarity cuts.}
\end{figure}

For example, consider the unitarity cuts of a one-loop four-point amplitude in
any gravity theory coming from the low energy limit of a string
theory.  As depicted in \fig{TwoParticleFigure}, a gravity tree
amplitude appears on each side of the cut.  Thus, on the cut one must
evaluate the product,
$$
\sum_{\rm gravity \atop states}
{\cal M}_4^{\rm tree} (1, 2, \ell_2, -\ell_1) \times 
     {\cal M}_4^{\rm tree} (3,4, \ell_1, -\ell_2) \,,
\equn
$$
summed over all states in the gravity theory that can cross the cut.
From the KLT relations (\ref{KLTExamples}) we may re-express this
product in terms of gauge theory amplitudes,
$$
\eqalign{
&\sum_{\rm gauge\atop states}
s_{12} \, A_4^{\rm tree} (1, 2, \ell_2, -\ell_1) \times
A_4^{\rm tree} (3,4, \ell_1, -\ell_2) \cr
\times \null &
\sum_{\rm gauge\atop states}
s_{12} \, A_4^{\rm tree} (1,2, \ell_1, -\ell_2) \times
A_4^{\rm tree} (3,4, -\ell_2,\ell_1) \,.\cr}
\equn
$$
The sum over the gravity theory states necessarily factorize into a
sum over the states of two gauge theories because the underlying 
string theory has this property.  This is a generic property 
holding for all cuts. 
In this way, one can re-express cut gravity
amplitudes in terms of gauge theory ones. This has led to the
construction of a number of non-trivial gravity loop
amplitudes~\cite{BDDPR,AllPlusGravity,MHVGravity}.


\section{Construction of a gravity Lagrangian from QCD}

We now discuss a procedure for constructing an off-shell low energy
Lagrangian for pure gravity starting from QCD gluon amplitudes.
We explicitly carry this out for up to five graviton interactions.  By
construction, the Lagrangian that we obtain is equivalent to the
Einstein-Hilbert Lagrangian in that it produces identical
tree-level $S$-matrix elements.

In field theory one usually takes a given Lagrangian and constructs
Feynman rules that can then be used to obtain the $S$-matrix elements.
Here we wish to reverse this process, since from the KLT relations we
have the gravity $S$-matrix in terms of QCD amplitudes, but wish
instead to obtain a Lagrangian that preserves the `left'-`right'
factorization of Lorentz indices. 

To obtain the $n$-graviton term in the Lagrangian we start with the
tree-level $n$-graviton amplitudes, given in terms of the QCD
amplitudes, and subtract all diagrams containing up to
$(n-1)$-graviton interaction vertices.  By iterating this procedure
we can systematically build a gravity Lagrangian.  However, to
jump-start this process we first need a propagator and three
vertex.

Consider the graviton propagator.  The standard de Donder gauge
propagator,
$$
P_{\mu\nu; \alpha\beta} (k) = \langle h_{\mu \alpha} h_{\nu\beta} \rangle_0
= {1 \over 2} 
\Bigl[\eta_{\mu\nu} \eta_{\alpha\beta} + \eta_{\mu\beta} \eta_{\nu\alpha} 
- {2\over D-2} \eta_{\mu\alpha} \eta_{\nu\beta} \Bigr] {i\over k^2 + i\eps}
\,, 
\equn\label{deDonderPropagator}
$$
is unacceptable since it contracts `left' and `right' indices; for example, 
the last term contracts $\mu$ with $\alpha$.  Moreover, it contains 
explicit dependence on the dimension $D$, which must somehow cancel from
the tree-level $S$-matrix elements since there is no such dependence in 
the KLT relations or in the gauge theory amplitudes.  
(We have chosen $\eta_{\mu\nu}$ to have signature $(+,-,-,-)$.)
A better propagator is~\cite{BDS}
$$
P_{\mu\nu; \alpha\beta} (k) = 
{i\over 2} 
{\eta_{\mu\nu} \eta_{\alpha\beta} + \eta_{\mu\beta} \eta_{\nu\alpha} 
 \over k^2 + i \eps}  \,.
\equn\label{SymmetricPropagator}
$$
However, even this propagator is unacceptable because of the second
term, which contracts a left index with a right one. To prevent
such contractions, we wish to use instead the propagator, 
$$
P_{\mu\nu; \alpha\beta} (k)  =
 \eta_{\mu\nu} \eta_{\alpha\beta}\, {i  \over k^2 + i \eps} \, .
\equn\label{NewPropagator}
$$
In a sense, for each graviton $h_{\mu\alpha}$ we assign the index
$\mu$ to be a `left' index and $\alpha$ to be a `right' index.  Of
course, since $h_{\mu\alpha}$ is symmetric it does not matter which
index is assigned to the left and which to the right, but once the
choice is made we demand that left indices never contract with right
ones.  At first sight the gravity propagator (\ref{NewPropagator})
might seem rather peculiar since it appears to violate the fundamental
property of the graviton being a symmetric tensor field.
Nevertheless, it is not difficult to show that one may use the
propagator~(\ref{NewPropagator}) instead
of~(\ref{SymmetricPropagator}) with no effect on the $n$-graviton
$S$-matrix elements, provided that all vertices satisfy a {\it rigid}
left--right interchange symmetry.  By a rigid symmetry we mean that
each $n$-point vertex be symmetric under a simultaneous interchange of
all left and right Lorentz indices,
$$
V_n^{\mu_1 \mu_2 \cdots \mu_n; \alpha_1 \alpha_2 \cdots \alpha_n} 
(k_1, k_2, \ldots , k_n)  
= V_n^{\alpha_1 \alpha_2 \cdots \alpha_n; \mu_1 \mu_2 \cdots \mu_n} 
(k_1, k_2, \ldots , k_n) \,,
\equn
$$
where the $\mu_i$ are the left indices associated with each 
graviton and the $\alpha_i$ are the right indices.

\begin{figure}[ht]
\centerline{\epsfxsize 1.2 truein \epsfbox{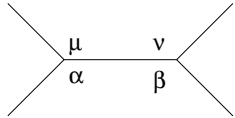}}
\vskip -.2 cm
\caption[]{
\label{PropagFigure}
\small A sample diagram for showing that we may use propagator 
(\ref{NewPropagator}) instead of (\ref{SymmetricPropagator}).}
\end{figure}

With this requirement, one can show that there is no need to
symmetrize the graviton propagator in the left and right indices.
Consider, for example, the diagram in \fig{PropagFigure}.  The
external legs of the diagrams do not need to be explicitly symmetrized
since the external graviton polarization tensors automatically symmetrize the
legs.  The interchange of the $\mu$ and $\alpha$ indices can be undone
by performing a rigid interchange of left and right indices on the
left-most vertex.  Although the rigid interchange will also flip the indices on
the external legs, this can be undone since the indices of each
external leg are contracted with a symmetric tensor polarization.  The
same type of argument works for general diagrams. Thus, the propagator
(\ref{NewPropagator}) can be used instead of the propagator
(\ref{SymmetricPropagator}) since the symmetrization of indices
is automatically taken care of by the rigidly symmetrized vertices 
and by the symmetric polarization tensors on the external legs.

%

Given the propagator (\ref{NewPropagator}) we must also choose a three
vertex.  Any three vertex which agrees on shell with the
three-graviton vertex is suitable for our purposes.  (Kinematic
restrictions prevent a massless particle from decaying into two
others, but in terms of formal polarization tensors the vertex does
not vanish; below we shall also obtain the same vertex from the
Einstein-Hilbert action.)  From string theory (see
e.g. ref.~\cite{GSW}), the on-shell three-graviton vertex can be
expressed in terms of products of gauge theory vertices.  This
motivates the choice of the three graviton vertex,
$$
\eqalign{
i G_3^{\mu\alpha, \nu\beta, \rho\gamma}(k, p , q) & = - {i \over 2} 
\Bigl( {\kappa \over 2} \Bigr)
\Bigl[
V_{\rm GN}^{\mu\nu\rho}(k, p, q) 
\times V_{\rm GN}^{\alpha \beta \gamma}(k, p, q) +
V_{\rm GN}^{\nu\mu\rho}(p, k, q) \times 
V_{\rm GN}^{\beta \alpha \gamma}(p, k, q) \Bigr]\,, \cr }
\equn\label{GravityYMThreeVertex}
$$
where
$$
 V^{\mu\nu\rho}_{\rm GN}(1,2,3) = 
       i\sqrt{2}  \bigl( k_1^\rho \, \eta^{\mu\nu}
            + k_2^\mu \eta^{\nu\rho} + k_3^\nu \eta^{\rho\mu} \bigr)\,, 
\equn
$$
is the color ordered Gervais-Neveu~\cite{GN} Yang-Mills three vertex,
from which the color factor has been stripped. Our main reasons for
using the vertex (\ref{GravityYMThreeVertex}) is its simplicity and
the fact that it makes the relationship to gauge theory manifest.  By
construction, this vertex is symmetric under the rigid interchange of
left and right indices.

\begin{figure}[ht]
\centerline{\epsfxsize 3.8 truein \epsfbox{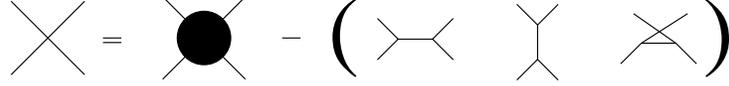}}
\vskip -.2 cm
\caption[]{
\label{BootStrapFigure}
\small One can obtain a four vertex with the left-right
factorization property by subtracting the diagrams containing the
kinematic poles from the $S$-matrix.}
\end{figure}

Armed with the propagator~(\ref{NewPropagator}) and the three vertex
(\ref{GravityYMThreeVertex}) we may then use the KLT relations to find
a gravity Lagrangian using QCD gluon amplitudes as input.  At the first step
of the process one defines a four-vertex by subtracting from the
four-point $S$-matrix obtained via the KLT relations all diagrams
containing a kinematic pole, as shown in \fig{BootStrapFigure}.  This
four vertex automatically has the property that left Lorentz indices
do not contract with right ones, since the gravity $S$-matrix and
diagrams containing the three-point vertex
(\ref{GravityYMThreeVertex}) have this property.  Moreover, the vertex
defined in this way also has the rigid symmetry under an interchange
of left and right indices.

%

We may then convert this four-vertex to an $h^4$ term in the
Lagrangian by inverting the usual procedure of obtaining Feynman
vertices from Lagrangians.  Of course, there is some ambiguity in this
process since one can always add terms which vanish on shell.  As long
as these terms do not mix left and right Lorentz indices and satisfy a
rigid left-right symmetry, which can always be imposed by hand, they
are acceptable terms; differences in the four-vertex then induce
differences in the deduced higher-point vertices.

Once the four-graviton terms in the Lagrangian have been chosen we can
then continue the process to obtain a five vertex by subtracting from
the five-graviton amplitude all diagrams containing three- and
four-point vertices.  In principle, one can continue in this way to an
arbitrary number of external legs, allowing one to build a Lagrangian
for gravity order by order in perturbation theory using QCD gluon
amplitudes.  By construction this Lagrangian has the property that it
generates all tree-level $S$-matrix elements and that it never
contracts left indices with right ones.  

Starting from QCD gluon tree amplitudes, we have carried out the
construction for up to five points yielding the local
gravity Lagrangian $L = \sum_i L_i$,
where
$$
\eqalign{
L_2 & = - {1\over 2} \, h_{\mu\nu} \partial^2 h_{\mu\nu}\,,  \cr
L_3 & = \kappa \Bigl[{1\over 2}h_{\mu\nu} h_{\rho\sigma,\mu\nu} h_{\rho\sigma}
          + h_{\nu\mu} h_{\rho\mu, \sigma} h_{\rho\sigma, \nu} \Bigr]\,, \cr
L_4 & = -\kappa^2 \Bigl[
         {1\over 32} h_{\mu\nu,\lambda}  h_{\mu\nu,\lambda} 
                         h_{\rho\sigma} h_{\rho\sigma}
        +{1\over 2} h_{\mu \nu,\lambda}h_{\mu\rho}h_{\sigma\nu,\rho}
                 h_{\sigma\lambda}
        +{1\over 8} h_{\mu \nu}h_{\mu \lambda,\rho}h_{\sigma \nu,\rho}
                 h_{\sigma \lambda} 
        -{1\over 4} h_{\mu\nu}h_{\mu\lambda}h_{\rho\sigma,\nu} 
                 h_{\rho\sigma,\lambda}\cr
& \hskip 1.3 cm 
        -{1\over 4} h_{\mu\nu} h_{\lambda\rho}h_{\lambda\sigma,\rho}
                 h_{\mu\sigma,\nu}
        +{1\over 16} h_{\mu\nu} h_{\mu\nu}h_{\lambda\rho} 
                 h_{\lambda\rho,\sigma\sigma} 
        +{1\over 24} h_{\mu\nu} h_{\lambda\nu} h_{\lambda\rho} 
                 h_{\mu\rho,\sigma\sigma}\Bigr] \,, \cr
L_5 & = \kappa^3 \Bigl[
         {3\over 64} h_{\mu\nu} h_{\mu\nu} h_{\rho\sigma} 
                 h_{\rho\sigma,\kappa\lambda} h_{\kappa\lambda} 
        -{1\over 12} h_{\mu\nu} h_{\mu\nu,\kappa} h_{\rho\sigma,\lambda}   
                 h_{\rho\kappa} h_{\lambda\sigma}
        -{1\over 48} h_{\mu\nu,\kappa} h_{\mu\nu} h_{\rho\sigma} 
                 h_{\rho\kappa,\lambda} h_{\lambda\sigma} \cr
&  \hskip 1.3 cm 
        -{1\over 12} h_{\mu\nu,\kappa} h_{\mu\nu,\lambda} h_{\rho\sigma} 
                 h_{\rho\kappa} h_{\lambda\sigma}
        -{3\over 32} h_{\mu\nu} h_{\mu\nu} h_{\rho\sigma,\kappa\lambda}  
                 h_{\rho\kappa} h_{\lambda\sigma} 
        -{1\over 12} h_{\mu\nu} h_{\rho\nu,\mu}  h_{\rho\sigma} 
                        h_{\lambda\sigma,\kappa} h_{\lambda\kappa} \cr
&  \hskip 1.3 cm 
        -{1\over 6}  h_{\mu\nu} h_{\rho\nu}  h_{\rho\sigma} 
                        h_{\lambda\sigma,\mu\kappa} h_{\lambda\kappa}
        +{1\over 6}  h_{\mu\nu} h_{\rho\nu,\kappa} h_{\rho\sigma,\mu} 
                        h_{\lambda\sigma} h_{\lambda\kappa}
        -{1\over 6}  h_{\mu\nu} h_{\rho\nu,\kappa} h_{\rho\sigma} 
                        h_{\lambda\sigma,\mu} h_{\lambda\kappa} \cr
&  \hskip 1.3 cm 
        +{1\over 12} h_{\mu\nu} h_{\rho\nu} h_{\rho\sigma,\mu\kappa} 
                        h_{\lambda\sigma} h_{\lambda\kappa}
        +{1\over 8} h_{\mu\nu} h_{\rho\nu,\lambda} h_{\rho\sigma} 
                        h_{\mu\sigma,\kappa} h_{\lambda\kappa} 
        +{1\over 24} h_{\mu\nu} h_{\rho\nu,\lambda\kappa} h_{\rho\sigma} 
                        h_{\mu\sigma} h_{\lambda\kappa} \Bigr]\,, \cr}
\equn
\label{GaugeLagrangian}
$$
where $h_{\mu\nu, \lambda} = \partial_\lambda h_{\mu\nu}$. 
In Minkowski space, one of any two contracted indices should be taken
to be a raised index using $\eta^{\mu\nu}$, but we have suppressed
this. In principle it might have been necessary to introduce
auxiliary fields for locality to hold; indeed, as discussed below, when
comparing this Lagrangian to the Einstein-Hilbert gravity Lagrangian
it is useful to introduce an auxiliary dilaton.  Although the
Lagrangian (\ref{GaugeLagrangian}) is not unique since the terms can
be modified by adding or subtracting contributions that vanish on
shell and appropriately modifying the higher-point contributions, it
is a relatively simple one.  More importantly, as we shall see below,
it allows for a relatively straightforward match to the conventional
Einstein-Hilbert Lagrangian.  By adjusting the four-point terms in the
Lagrangian we have found a solution for $L_5$ that contains only six
terms, but then the connection to the Einstein-Hilbert Lagrangian is a
bit more complicated.

For compactness we have removed the rigid symmetrization between left
and right indices; however, if one uses the propagator
(\ref{NewPropagator}), each vertex must be rigidly symmetrized between
left and right indices, e.g.,
$$
L_3  \rightarrow {\kappa\over 2}  
           \Bigl[h_{\mu\nu} h_{\rho\sigma, \mu\nu} h_{\rho\sigma}
          + h_{\nu\mu} h_{\rho\mu, \sigma} h_{\rho\sigma, \nu}
          + h_{\mu\nu} h_{\mu\rho, \sigma} h_{\sigma\rho, \nu} \Bigr]\,.
\equn
$$
Although we do not explicitly give the generated Feynman rules here,
it is quite straightforward to obtain these.  The advantages of using
Feynman rules generated by this Lagrangian instead of the conventional
de Donder gauge rule are clear: besides the fact that one can use the
propagator (\ref{NewPropagator}) instead of the more complicated de
Donder gauge propagator (\ref{deDonderPropagator}) the three-, four-
and five-point vertices are quite a bit simpler than the corresponding
de Donder gauge vertices.  Once the interaction terms have been
rigidly symmetrized, when deriving the Feynman rules we can in a sense
treat $h_{\mu\nu}$ as a tensor field with no special index symmetry.
Note that the Feynman diagrams generated by this Lagrangian do not
contain explicit factors of $D$.  In the conventional de Donder gauge
such factors do appear, but somehow cancel from the $S$-matrix
elements.

The relative simplicity of these Feynman rules is related to the
preservation of the $S$ matrix property that left and right Lorentz
indices should not contract with each other.  Although the KLT
relations might appear to suggest that field variables exist where gravity
can be reformulated as a polynomial theory with no more than
four-point interactions as in the gauge theory case, we
have been unsuccessful in finding such a Lagrangian.

The Lagrangian (\ref{GaugeLagrangian}) does contain general coordinate
invariance although it is not manifest since the symmetry has been
gauge fixed.  To make this explicit, we now relate the Lagrangian
(\ref{GaugeLagrangian}) to the usual Einstein-Hilbert one.  In order
to do so we must first eliminate all terms in the Einstein-Hilbert
Lagrangian which contract left with right graviton indices, i.e,
$$
h_{\mu\mu}\, , \hskip 1 cm 
h_{\mu\nu} h_{\nu\lambda} h_{\lambda\mu} \,, \hskip 1 cm 
\cdots, \hskip 1 cm \Tr[h^{2m+1}] \,,
\equn\label{BadTerms}
$$
where $\Tr[h^n] \equiv h_{\mu_1 \mu_2} h_{\mu_2 \mu_3} \cdots h_{\mu_n \mu_1}$.
Because of the way that these types of terms are tangled in the
Einstein-Hilbert Lagrangian, it is not obvious how one can accomplish
this.  Nevertheless, the existence of the Lagrangian
(\ref{GaugeLagrangian}) implies that there must be some rearrangement
with the desired property.

In ref.~\cite{BDS} the dilaton was introduced to
allow for a field redefinition which removes the graviton trace from
the quadratic terms in the Lagrangian.  The appearance of the dilaton
as an auxiliary field to help rearrange the Lagrangian is motivated by
string theory which requires the presence of such a field.  Following
the discussion of ref.~\cite{BDS}, we then consider the Lagrangian for
gravity coupled to a dilaton,
$$
L^{\rm EH} =   {2\over \kappa^2} \sqrt{-g} \, 
       R +  \sqrt{-g} \, \partial^\mu\phi\partial_\mu\phi \,.
\equn
$$
(Our conventions are those of Weinberg~\cite{Weinberg}, except that
we have flipped the signature of the metric, $g_{\mu\nu} \rightarrow
- g_{\mu\nu}$; this then induces an overall sign flip in the Lagrangian.)
In de Donder gauge, for example, taking $g_{\mu\nu} = \eta_{\mu\nu} +
\kappa h_{\mu\nu}$, the quadratic part of the Lagrangian is
$$
L_2 = 
- \frac{1}{2} h_{\mu\nu} \partial^2 h_{\mu\nu} 
+ \frac{1}{4} h_{\mu\mu} \partial^2 h_{\nu\nu}
-  \phi \partial^2 \phi \,.
\equn
$$
The term involving $h_{\mu\mu}$ can be eliminated with the
simultaneous field redefinitions~\cite{BDS},
$$
h_{\mu\nu} \rightarrow h_{\mu\nu} 
        + \eta_{\mu\nu}{\sqrt{\frac{2}{D-2}}}\, \phi\,,
\hskip 2 cm
\phi \rightarrow \frac{1}{2} h_{\mu\mu} + \sqrt{\frac{D-2}{2}}\, \phi\,,
\equn\label{FieldRedef}
$$
so that the Lagrangian reduces to
$$
L_2 \rightarrow 
- \frac{1}{2} h_{\mu\nu} \partial^2 h_{\mu\nu} 
+  \phi \partial^2 \phi\,.
\equn\label{TwoLagrange}
$$
(This field redefinition is a bit different than the one in ref.~\cite{BDS}
because we have chosen to normalize the dilaton kinetic term differently
so as to slightly simplify the induced interaction terms discussed below.)

For the case of purely graviton external states the dilaton will not
contribute to the tree-level $S$-matrix because of the selection rule
that $\phi$ must be created or annihilated in pairs.  Of course, if
the dilaton, or any matter fields
appearing in the theory, are taken as external states they can
propagate in intermediate states.  In this case, one would need to
specify the precise matter content of the theory before proceeding
with the analysis.  Nevertheless, the fundamental factorization of the
$S$-matrix implied by the KLT relations would still hold, since it
follows from the underlying string theory.

Here we wish to generalize the field redefinitions (\ref{FieldRedef})
to all orders in $\kappa$ so as to remove all terms of the form
(\ref{BadTerms}) from the Lagrangian.  Our generalization for the case of
no gauge fixing is
$$
g_{\mu\nu} = e^{\sqrt{\frac{2}{D-2}} \kappa \phi} 
e^{\kappa\, h_{\mu\nu}}  \equiv 
 e^{\sqrt{\frac{2}{D-2}} \,\kappa \phi} 
\Bigl(\eta_{\mu\nu} + \kappa h_{\mu\nu} + {\textstyle \kappa^2\over 2} 
                    h_{\mu\rho} h_{\rho\nu} + \cdots \Bigr) \,,
\equn\label{StringVars}
$$
where $h_{\mu\nu}$ is the graviton field, 
followed by the change of variables
$$
\phi \rightarrow -\sqrt\frac{2}{D-2} \, 
\Bigl( \phi + \frac{1}{2} h_{\rho\rho}\Bigr) \,.
\equn\label{PhiRedef}
$$
We have verified through $\Ord(h^6)$ that this choice eliminates all
terms (\ref{BadTerms}) which mix left and right Lorentz indices, 
even before fixing the gauge.
As yet we have not performed
any gauge fixing so the action is generally coordinate invariant, even
if the choice of field variables obscures this.

One might be concerned that the field redefinition (\ref{StringVars})
would alter the gravity $S$-matrix.  However, the $S$-matrix is
guaranteed to be invariant under non-linear field redefinitions or
under linear ones that do not alter the coupling to the external
traceless polarization tensors.  Indeed, our explicit calculations
respect this property, as required.

An important remaining question is how one can choose a gauge fixing
so that the terms in the Einstein-Hilbert Lagrangian resemble the
terms of the Lagrangian (\ref{GaugeLagrangian}) deduced from the QCD
amplitudes.  We have found a solution, which is to replace the field
redefinition (\ref{PhiRedef}) with a non-linear generalization,
$$
\phi \rightarrow - \sqrt\frac{2}{D-2}\biggl[ 
\biggl( \phi + \frac{1}{2} {\rm Tr}\, h \biggr) +
\kappa \biggl( \frac{1}{4} \phi^2 - \frac{1}{8} {\rm Tr}\,(h^2) \biggr)+
\kappa^2 \biggl( \frac{1}{12} \phi^3 - \frac{1}{8} \phi\, {\rm Tr}\,(h^2)
\biggr) +\cdots\biggr] \,.
\equn\label{PhiRedefB}
$$
Then we add a gauge fixing term to the Lagrangian, $(F_\mu)^2$ following
the standard procedure, where 
$$
\eqalign{
F_\mu & = \biggl( h_{\mu\nu,\nu} + \phi,_\mu \biggr)
+ \kappa \biggl( -\frac{1}{4} {\rm Tr}\,(h^2)_{,\mu}
        -\frac{1}{2} \phi h_{\mu\nu,\nu} - h_{\mu\nu}  \phi,_\nu \biggr) 
+ \kappa^2 \biggl( \frac{1}{16} {\rm Tr}\,(h^2) h_{\mu\nu,\nu} 
+\frac{1}{8} {\rm Tr}\,(h^2)_{,\nu} h_{\mu\nu}\cr
&  \hskip 3 cm 
-\frac{1}{12} h_{\mu\nu} h_{\lambda\nu,\rho} h_{\lambda\rho}
+\frac{1}{6} h_{\mu\nu,\rho} h_{\lambda\nu} h_{\lambda\rho}
+\frac{1}{24} h_{\mu\nu} h_{\lambda\nu} h_{\lambda\rho,\rho}
- \frac{1}{8} ( \phi\,{\rm Tr}\,(h^2) )_{,\mu} \biggr) +\cdots\,. \cr}
\equn
$$
With these choices we then obtain the desired form of the Lagrangian
through $\Ord(h^4)$, which we express in terms of the Lagrangian
(\ref{GaugeLagrangian}) derived from QCD amplitudes,
$$
\eqalign{
L_2^{\rm EH} & = L_2 + \phi \partial^2 \phi \,, \cr
L_3^{\rm EH} & = L_3 -
      {\kappa\over 2} h_{\mu\nu,\kappa} h_{\mu\nu,\kappa} \phi \,, \cr
L_4^{\rm EH} & = L_4  
+ \kappa^2 \Bigl[ {1\over 32} h_{\mu\nu,\lambda} h_{\mu\nu,\lambda} 
      h_{\rho\sigma} h_{\rho\sigma}
+{1\over 2} \phi h_{\mu\nu} h_{\lambda\sigma,\mu} h_{\lambda\sigma,\nu}
-{1\over 4}\phi_{,\mu} h_{\mu\nu} h_{\lambda\nu} h_{\lambda\sigma,\sigma}
+{1\over 2}\phi_{,\mu} h_{\mu\nu}  h_{\lambda\sigma,\nu} h_{\lambda\sigma} \cr
& \hskip 1.5 cm 
-{1\over 2}\phi_{,\mu} h_{\mu\nu}  h_{\lambda\nu,\sigma} h_{\lambda\sigma}
-\phi h_{\mu\nu} h_{\mu\lambda,\sigma} h_{\sigma\lambda,\nu}
-{1\over 8} \phi_{,\mu} h_{\mu\rho,\rho} h_{\lambda\sigma} h_{\lambda\sigma}
+{1\over 8} \phi^2 h_{\mu\nu,\lambda} h_{\mu\nu,\lambda} \Bigr] \,.\cr}
\equn
$$

For the case of pure graviton external states we may integrate out
$\phi$ from the path integral.  Equivalently, we may substitute the 
equation of motion for $\phi$,
$$
\phi = {\kappa\over 4} \, {1\over \partial^2}  
         (h_{\mu\nu,\kappa} h_{\mu\nu,\kappa}) + \cdots
\equn
$$
into the Lagrangian.  For the two- and three-graviton terms in the
Lagrangian, this gives exactly the terms in the Lagrangian
(\ref{GaugeLagrangian}).  For the four-point we obtain exactly $L_4$ 
plus a non-local piece that vanishes for on-shell gravitons,
$$
\Delta L_4 = {\kappa^2 \over 16} h_{\mu\nu,\kappa} h_{\mu\nu,\kappa}
\, {1\over \partial^2} \, (h_{\rho\sigma, \lambda\lambda} h_{\rho\sigma})
\equn
$$
Since the four-point Lagrangians differ a bit off-shell,
at the five-point level, terms that are non-zero on-shell need to
be added to $L_5$,
$$
\Delta L_5\Bigr|_{\rm on\ shell} = \kappa^3 \Bigl[
     {1\over 64} h_{\mu\nu} h_{\mu\nu} h_{\rho\sigma} 
                 h_{\rho\sigma,\kappa\lambda} h_{\kappa\lambda} 
        -{1\over 16} h_{\mu\nu,\kappa} h_{\mu\nu} h_{\rho\sigma} 
                 h_{\rho\kappa,\lambda} h_{\lambda\sigma} 
        -{1\over 32} h_{\mu\nu} h_{\mu\nu} h_{\rho\sigma,\kappa\lambda}  
                 h_{\rho\kappa} h_{\lambda\sigma} \Bigr] \,.
\equn
$$

In a sense we may take $L^{\rm EH}$ as the more fundamental one since
it comes directly from the Einstein-Hilbert Lagrangian, while the
Lagrangian in \eqn{GaugeLagrangian} provides the guidance needed to
obtain it. This shows the connection of the Lagrangian obtained from
QCD (\ref{GaugeLagrangian}) and the Einstein-Hilbert Lagrangian for a
very particular set of field variables and and gauge fixing.
Although the two Lagrangians differ somewhat even after
integrating out the auxiliary dilaton, we have explicitly shown that
for up to five gravitons they generate the same on-shell scattering
amplitudes.  

Presumably, similar results can be obtained without introducing the
dilaton; nevertheless, we found it useful for clarifying the required
reorganization of the Einstein-Hilbert Lagrangian.  
Other reorganizations based on introducing 
other fields such as an anti-symmetric tensor, which
is also motivated by string theory, are also possible~\cite{Siegel}.

At loop level there are also ghost contributions that can be
obtained via the usual methods. (The Jacobian generated by the field
redefinition is trivial, at least in perturbation theory, since it is
a point transformation.)  Additionally the auxiliary dilaton would
propagate in the loops, which would then need to be subtracted (see
e.g., ref.~\cite{BDS}).  At loop level the unitarity method advocated
in refs.~\cite{Review,BDDPR,AllPlusGravity,MHVGravity} is, however, an
efficient way to obtain the $S$-matrix without the need for Feynman
rules.

\section{Sample applications}

As one simple application, one may obtain the soft factor for gravity
amplitudes directly from the soft behavior of QCD amplitudes. Gravity
tree amplitudes have the well known behavior~\cite{WeinbergSoftG},
$$
{\cal M}_n^\tree(1,2,\ldots,n^+)\ \mathop{\longrightarrow}^{k_n\to0}\
   {\kappa\over2} \Soft^{\rm gravity}(n^+) \times\ 
   {\cal M}_{n-1}^\tree(1,2,\ldots,n-1) \,,
\equn\label{GravTreeSoft}
$$
as the momentum of graviton $n$, which we have taken to carry positive 
helicity, becomes soft. Using the three-graviton vertex
(\ref{GravityYMThreeVertex}) which is expressed in terms of the QCD 
three-gluon vertex, 
the gravity soft factor can then be expressed in terms of the QCD soft factor:
$$
\Soft^{\rm gravity}(n^+) = \sum_{i=1}^{n-1} s_{ni} 
            \Soft^{\rm QCD}(q_l, n^+, i) \times 
            \Soft^{\rm QCD}(q_r, n^+, i) \,,
\equn\label{FinalSoftGrav}
$$
where $\Soft^{\rm QCD}(q, n^+, i) = \spa{q}.i/\spa{q}.{n}\spa{n}.i$
is the eikonal factor for a positive helicity soft gluon in QCD.  The
$\spa{i}.j= \la i^- | j^+\ra$ are spinor inner products and
$|i^{\pm}\ra$ are massless Weyl spinors of momentum $k_i$, labeled
with the sign of the helicity (see e.g.,~\cite{ManganoReview}).
Although not manifest, the soft factor (\ref{FinalSoftGrav}) is
independent of the choices of null helicity `reference momenta' $q_l$ and
$q_r$.  By choosing $q_l = k_1$ and $q_r = k_{n-1}$ we recover the
form of the soft graviton factor for $k_n \rightarrow 0$ used
in, for example, refs.~\cite{BGK,AllPlusGravity,MHVGravity}.

As a less trivial example, the explicit factorization of the left and
right indices of the interaction vertices would imply
that the all-plus helicity graviton current satisfying
eqs.~(B.10) and (B.11) of ref.~\cite{MHVGravity} do actually
follow from Einstein gravity.  In ref.~\cite{MHVGravity} the hitch in
obtaining these recursion relations from Einstein gravity was the
different choices of helicity reference momenta on the left and on the
right. (Although, not derived directly from Einstein gravity these
recursion relations were useful for defining `half-soft' functions
which serve as building blocks for one-loop gravity amplitudes.)  In
this case, $\pol_+^{\mu\alpha} \pol_+^{\beta\mu} \not = 0$, which
prevented a derivation of the recursion relations from Einstein
gravity.  With the factorization of left and right Lorentz indices, such
contractions simply do not occur.


\section{Discussion}

Since the tree-level $S$-matrix is generated by solving perturbatively
the classical equations of motion, one might suspect that it is
possible to relate more general solutions of the classical equations
of motion for gravity to gauge theory solutions.  The property that
the $S$-matrix can be expressed in terms of `left' and `right' sectors
is generic in string theory.  For this reason, it should be possible
to extend the discussion in this letter to theories containing, for
example, the anti-symmetric tensor or to supergravity theories.  It
also seems reasonable that gravity theories in curved spaces 
can be reformulated to express tree-level
amplitudes in terms of gauge theory ones.  Moreover,
the intimate connection of perturbative gravity amplitudes with gauge
theory ones suggests a closer connection of diffeomorphism invariance
with non-abelian gauge symmetry than one might have suspected.  We
feel that these issues deserve further attention.

\vskip .2 cm 

We thank S. Cherkis, L. Dixon and J. Schwarz for helpful discussions.


\end{document}